\documentclass[aps,prb,twocolumn,amsmath,amssymb,nofootinbib,eqsecnum,superscriptaddress,floatfix,preprintnumbers]{revtex4}

\usepackage{amsmath}
\usepackage{amssymb}
\usepackage{amsthm}
\usepackage[dvips]{color} 
\usepackage{graphicx}
\usepackage{dcolumn} 
\usepackage{bm} 
\usepackage[hypertex]{hyperref}
\usepackage{longtable}
\usepackage{ulem}   
\newcommand{\bs}[1]{{\boldsymbol{#1}}}

\begin{document}

\author{Claudio Chamon} 
\affiliation{
Physics Department, 
Boston University, 
Boston, Massachusetts 02215, USA
            } 

\author{Roman Jackiw} 
\affiliation{
Physics Department, 
Massachusetts Institute of Technology, 
Cambridge, Massachusetts 02139, USA
            }

\author{So-Young Pi} 
\affiliation{
Physics Department, 
Boston University, 
Boston, Massachusetts 02215, USA
            } 

\author{Luiz Santos} 
\affiliation{
Physics Department, 
Harvard University, 
Cambridge, Massachusetts 02138, USA
            }

\preprint{MIT-CTP/4270}

\date{\today}

\title{Conformal quantum mechanics as the CFT$_1$ dual to AdS$_2$ }

\begin{abstract}
  A $0+1$-dimensional candidate theory for the CFT$_1$ dual to
  AdS$_2$ is discussed. The quantum mechanical system does not have a
  ground state that is invariant under the three generators of the
  conformal group. Nevertheless, we show that there are operators in
  the theory that are not primary, but whose ``non-primary character''
  conspires with the ``non-invariance of the vacuum'' to give
  precisely the correlation functions in a conformally invariant
  theory.
\end{abstract}
\maketitle

\section{
Introduction
        }
\label{sec: Introduction}

An elementary realization of the AdS/CFT 
correspondence~\cite{Maldacena97} proceeds as follows. Consider a 
scalar field $\Phi$ on a ($d + 1$)-dimensional AdS space
[in Poincar\'e coordinates ($z$,$x^{i}$), $i = 1,...,d$, 
with boundary at $z=0$]. The field is governed by the action
$I(\Phi)$, which leads to equations of motion for $\Phi$ on the
background AdS space. When the equations are solved, subject to the
boundary condition $\Phi(z,\bs{x}) \rightarrow \phi(\bs{x})$ as $z \rightarrow 0$,
the solution provides us with a functional of (the unspecified)
$\phi$: $\Phi(\phi)$. The action, evaluated on this particular
solution results in a further functional of 
$\phi$: $I(\Phi)\big|_{\Phi = \Phi(\phi)} \equiv W(\phi)$.
In the AdS/CFT correspondence the functional  $W(\phi)$ is 
identified with the generating functional in $d$ dimensions
for the $n$-point correlation functions of the operators 
$O(\bs{x})$ sourced by $\phi(\bs{x})$.
\begin{equation}
\label{eq: AdS correlation function}
\langle
O(\bs{x}_{1})\,\cdots\,O(\bs{x}_{n})
\rangle
=
\frac{\delta}{\delta\,\phi(\bs{x}_{1})}\,
\cdots\,
\frac{\delta}{\delta\,\phi(\bs{x}_{n})}\,
W(\phi)\,\big|_{\phi = 0}
\end{equation}
The form of these operators and the theory governing them
remain unknown. But the $d$-dimensional dynamics and the
averaging state $\langle\cdots\rangle$ are taken to be
conformally invariant. (Boldface coordinates refer to $d$ 
dimensional space-time.)

In a simple application of this procedure one finds a $2$-point
function~\cite{Freedman},
\begin{equation}
\label{eq: 2 pt functions}
G_{2}(\bs{x},\bs{y})
\equiv
\langle
O(\bs{x})\,O(\bs{y})
\rangle\propto
\,
\frac{1}{|\bs{x}-\bs{y}|^{2\Delta}}	
\end{equation}
and a $3$-point function~\cite{Freedman},
\begin{eqnarray}
\label{eq: 3 pt functions}
&&
G_{3}(\bs{w},\bs{x},\bs{y})
\equiv
\langle
O_{1}(\bs{w})\,O_{2}(\bs{x})\,O_{3}(\bs{y})
\rangle \propto
\\
&&
\frac{1}
{
|\bs{w}-\bs{x}|^{\Delta_{1}+\Delta_{2}-\Delta_{3}}
|\bs{x}-\bs{y}|^{\Delta_{2}+\Delta_{3}-\Delta_{1}}
|\bs{y}-\bs{w}|^{\Delta_{3}+\Delta_{1}-\Delta_{2}}}
\nonumber
\;.	
\end{eqnarray}
These expressions are consistent with the putative
conformal invariance, where the operators $O_{i}$
are conformal primaries carrying dimension $\Delta_{i}$,
while the state in which the correlations are taken is a 
conformally invariant ``vacuum''.

The above development can be carried out for any
dimension, but ``the best understood...of AdS/CFT dualities
is the case AdS$_{3}$/CFT$_{2}$ largely because the conformal
group is infinite-dimensional (in two dimensions) and greatly 
constrains the dynamics... In lower dimensions - namely the
AdS$_{2}$ case...very little is understood.''~\cite{Strominger} 
Our goal is to describe in greater detail some features of the
AdS$_{2}$/CFT$_{1}$ duality.
In string theory AdS$_{2}$/CFT$_{1}$ is interesting because all
known black holes have an AdS$_{2}$ factor in their horizon 
geometry (AdS$_{2}$$\times$K). 
However, in our investigation 
the AdS$_{2}$ geometry stands alone and no reference is made
to strings or black holes.

Here we specifically inquire whether the 
results~(\ref{eq: 2 pt functions}) and~(\ref{eq: 3 pt functions}) 
for the $2$- and $3$-point functions can arise in a 
conformal quantum theory defined on a $1$-dimensional base space,
i.e. time. Thus we work with quantum mechanics of a particle on a
half-line subject to an inverse square interaction potential.
The scale invariance of this model was identified in Ref.~\onlinecite{Jackiw72},
and its properties were thoroughly analyzed by de Alfaro, Fubini and Furlan
(dAFF).~\cite{Fubini} [Other conformally (=SO($2,1$)) invariant quantum mechanical models
involve multicomponent variables~\cite{Strominger-2},
singular potentials~\cite{Jackiw-Son} and/or various magnetic velocity-dependent
interactions~\cite{Jackiw-3}, but they offer no further insights.] Our arguments
rely on the underlying SO($2,1$) group structure, not on the specific dynamics.

A challenge that we face in studying the conformally invariant
quantum mechanics is that in its Hilbert space there is no invariant vacuum state that 
is annihilated by all the generators of the $SO(2,1)$ group. We show, however,
that this does not pose an obstacle to defining correlation
functions of the form~(\ref{eq: 2 pt functions}) and~(\ref{eq: 3 pt functions}),
provided one identifies the relevant state and operators in the correlation functions.
We shall present two equivalent formulations of such states and operators,
which give rise to correlation functions obeying the constraints of conformal symmetry.

This paper is organized as follows. In Sec.~(\ref{sec: Review}) we 
review the symmetry  properties of AdS$_{2}$ and CFT$_{1}$ and introduce
the conformal invariant quantum mechanical model possessing $SO(2,1)$ symmetry studied by dAFF.  
In Sec.~(\ref{sec: Puzzle}) we 
discuss how states and operators, though not transforming according
to the conformal symmetry, combine to give rise to conformally
invariant correlation functions. In Sec.~(\ref{sec: Realizing an operator-state correspondence with
neither an invariant vacuum nor a primary operator}) we formulate an operator-state
correspondence for the CFT$_{1}$ and discuss how it can account for
correlation functions with conformal scaling behavior. Details of some calculations
are presented in the Appendix.

\section{
Review
        }
\label{sec: Review}

We begin by reviewing needed formulas. The AdS$_{2}$ (Euclidean)
line interval reads
\begin{equation}
ds^{2} = \frac{1}{z^2}\,(dz^2+dt^2).	
\end{equation}
(The $d$-dimensional ``$\bs{x}$'' collapses to the 
$1$-dimensional ``$t$''.) Killing vectors are 
conveniently presented with complex coordinates
$x = t + \mathrm{i} z$.
\begin{equation}
k^{(n)}
=
x^{(n-1)}\,\frac{\partial}{\partial\,x}
+
(x^{*})^{(n-1)}\,\frac{\partial}{\partial\,x^{*}},
\quad	n=1,2,3
\end{equation}
They satisfy the $SO(2,1)$ algebra.
\begin{equation}
\left[
k^{(m)},k^{(n)}
\right]	
=
(n-m)\,k^{(m+n-2)}
\end{equation}
This same algebra can be canonically realized in 
conformal quantum mechanics with operators 
$H$, $D$ and $K$, which also follow the $SO(2,1)$ commutators,
\begin{eqnarray}
\label{eq: commutation relations SO(2,1)}
&&
\mathrm{i}\,
\left[
D,H
\right]
=
H
\nonumber\\
&&
\mathrm{i}\,
\left[
D,K
\right]
=
-K
\nonumber\\
&&
\mathrm{i}\,
\left[
K,H
\right]
=
2D
\end{eqnarray}
or in the Cartan basis
\begin{eqnarray}
\label{eq: Cartan basis}
&&
R\equiv
\frac{1}{2}
(\frac{K}{a}+a\,H),
\quad
L_{\pm}
\equiv
\frac{1}{2}
(\frac{K}{a}-a\,H)
\pm \mathrm{i}\,D
\nonumber\\
&&
\left[
R,L_{\pm}
\right]
=
\pm L_{\pm},
\quad
\left[
L_{-},L_{+}
\right]
=
2R.
\end{eqnarray}
(The parameter ``$a$'', with time dimensionality,
is introduced for dimensional balance.) The coincidence between the
$SO(2,1)$ isometry of AdS$_{2}$ and the $SO(2,1)$ symmetry of a conformal quantum system 
is the basis of the AdS$_{2}$/CFT$_{1}$ correspondence
($R\,\sim\,k^{(2)}$, $L_{+}\,\sim\,k^{(3)}$, $L_{-}\,\sim\,k^{(1)}$).

$R$ can be taken to be a positive operator. It generates a compact subgroup.
According to representation theory for $SO(2,1)$ the spectrum of $R$ is
discrete.
\begin{eqnarray}
\label{eq: spectrum of R}
&&
R\,|n\rangle
=
r_{n}\,|n\rangle,
\nonumber\\
&&
r_{n} = r_{0} + n,
\quad
n=0,1,\cdots, 
\quad r_{0}>0
\nonumber\\
&&
\langle\,n'|n\,\rangle = \delta_{n',n}	
\end{eqnarray}
Ladder operators $ L_{\pm}$ act as
\begin{equation}
\label{eq: ladder operators}
L_{\pm}\,|n\rangle
=
\sqrt{r_{n}(r_{n} \pm 1) - r_{0}(r_{0}-1)}\,
|n \pm 1\,\rangle.
\end{equation}
Eq.(\ref{eq: ladder operators}) implies that
\begin{equation}
\label{eq: eigenstates of R}
|n\rangle
=
\sqrt{\frac{\Gamma(2r_{0})}{n!\,\Gamma(2r_{0}+n)}}\,
(L_{+})^{n}\,|0\rangle.	
\end{equation}
The $r_{0}$ eigenvalue of the lowest state
- the $R$ ``vacuum'' $|0\rangle$ - is connected to the Casimir invariant $\mathcal{C}$.
\begin{eqnarray}
\label{eq: Casimir}
&&
\mathcal{C}
\equiv
\frac{1}{2}\,
(H\,K + K\,H) - D^{2}
=
R^{2}-L_{+}\,L_{-}
\nonumber\\
&&
\mathcal{C}\,|n\rangle=
r_{0}(r_{0}-1)|n\rangle	
\end{eqnarray}
The above $SO(2,1)$ structure is
realized by dAFF in a canonical model, with
\begin{eqnarray}
\label{eq: dFF model}
&&
H = \frac{1}{2}(p^2 + \frac{g}{q^2}),
\quad g>0
\nonumber\\
&&
D = t\,H - \frac{1}{4}(pq + qp)
\nonumber\\
&&
K = -t^{2}\,H + 2\,t\,D + \frac{1}{2}\,q^2
\nonumber\\
&&
\mathrm{i}
\left[
p(t),q(t)
\right]	 = 1
\nonumber\\
&&
\mathcal{C} = \frac{g}{4}-\frac{3}{16}
\nonumber\\
&&
r_{0}
=
\frac{1}{2}\,(1+\sqrt{g+\frac{1}{4}}).
\end{eqnarray}
$H$, $D$ and $K$ are time-independent; q has
scale dimension $-1/2$ and is a conformal primary.
\begin{eqnarray}
\label{eq: q(t) transformations}
&&
\mathrm{i}
\left[
H,q(t)
\right]	 = \frac{d}{d\,t}\,q(t)
\nonumber\\
&&
\mathrm{i}
\left[
D,q(t)
\right]	 = t\frac{d}{d\,t}\,q(t) - \frac{1}{2}\,q(t)
\nonumber\\
&&
\mathrm{i}
\left[
K,q(t)
\right]	 = t^{2}\,\frac{d}{d\,t}\,q(t) - t\,q(t)
\end{eqnarray}
In fact we shall mainly utilize the group structure 
summarized in
(\ref{eq: commutation relations SO(2,1)})-(\ref{eq: Casimir}). 
The specific realization (\ref{eq: dFF model}) and (\ref{eq: q(t) transformations})
plays a secondary role. 

\section{Puzzle}
\label{sec: Puzzle}

Since dAFF present an explicit $SO(2,1)$-invariant CFT$_{1}$ model,
we inquire whether states and operators in that model reproduce
the conformally invariant correlation functions determined by
the AdS$_{2}$ correspondence.

Now we can state our puzzle about the AdS$_{2}$/CFT$_{1}$ duality.
In the dAFF model $SO(2,1)$ invariant states are not normalizable,
so forming diagonal matrix elements is problematical. 
Moreover, no state is invariant under all three $SO(2,1)$ transformations.
On the other hand,
normalizable, non-invariant states interfere with derivations of
conformal constraints. Furthermore, the AdS$_{2}$ calculation
indicates that the averaged operators carry (unspecified) arbitrary
dimensions, while the canonical model involves operators with fixed
rational dimensions.

Nevertheless it is intriguing that dAFF find amplitudes that match
precisely the forms found in the AdS$_{2}$ calculation. These are
constructed in dAFF as follows. States ``$\langle t |$'' are 
introduced on which the action of the $SO(2,1)$ generators is
realized with `` t ''- derivation.
\begin{eqnarray}
\label{eq: t representation}
&&
\langle t |\,H =\mathrm{i}\,\frac{d}{d\,t}\,\langle t |,
\nonumber\\
&&
\langle t |\,D =\mathrm{i}\,(t\frac{d}{d\,t}+r_{0})\,\langle t |
\nonumber\\
&&
\langle t |\,K =\mathrm{i}\,(t^{2}\frac{d}{d\,t}+2\,r_{0}\,t)\,\langle t |	
\end{eqnarray}
These ``t''-based generators satisfy the Lie algebra (\ref{eq: commutation relations SO(2,1)}) and lead to the
correct Casimir (\ref{eq: Casimir}). Explicit formulas are
found for $\langle t | n \rangle$ by solving the equation
\begin{eqnarray}
&&
\langle t| R | n \rangle = r_{n}\,\langle t | n \rangle
\nonumber\\
&=&
\frac{\mathrm{i}}{2}\,
\left[
(a+\frac{t^2}{a})\frac{d}{d\,t}+2\,r_{0}\,\frac{t}{a})
\right]\,\langle t | n \rangle
\end{eqnarray}
\begin{eqnarray}
\label{eq: beta definition}
&&
\langle t | n \rangle \equiv \beta_{n}(t) =
\\
&&
(-1)^{n}\,
\left[
\frac{\Gamma(2r_{0}+n)}{n!}
\right]^{1/2}\,
\left(
\frac{a-\mathrm{i}t}{a+\mathrm{i}t}
\right)^{r_{n}}\,
\frac{1}{\left( 1 + \frac{t^2}{a^2}  \right)^{r_{0}}}.
\nonumber
\end{eqnarray}
With these dAFF find
\begin{eqnarray}
\label{eq: 2-pt function dFF}
F_{2}(t_{1},t_{2})
&=&
\sum_{n}\,\beta_{n}(t_{1})\beta^{*}_{n}(t_{2})
\nonumber\\
&\equiv& \langle t_{1} | t_{2} \rangle \propto
\frac{1}{|t_{1}-t_{2}|^{2\,r_{0}}}
\end{eqnarray}
\begin{eqnarray}
\label{eq: 3-pt function dFF}
F_{3}(t_{1}, t , t_{2})
&\equiv& \langle t_{1} | B(t) | t_{2} \rangle
\nonumber\\
&\propto&
\frac{1}
{
|t-t_{1}|^{\delta}
\,
|t_{2}-t|^{\delta}
\,
|t_{1}-t_{2}|^{-\delta + 2r_{0}}
}.
\end{eqnarray}
Here $B$ is an unspecified primary with dimension $\delta$.

These correlators are precisely of the form
(\ref{eq: 2 pt functions}) and (\ref{eq: 3 pt functions}),
obtained in the AdS$_{2}$ calculation. Upon comparing with 
(\ref{eq: 2-pt function dFF}) and (\ref{eq: 3-pt function dFF}), we see that $G_{2}\,\sim\,F_{2}$ is 
the expected value of two operators, each with the effective
dimension $r_{0}$, while $G_{3}\,\sim\,F_{3}$ involves these
same two operators and a third operator $B$ with dimension $\delta$.

To resolve our puzzlement it remains to identify within CFT$_{1}$ the
averaging states and the two operators that form $G_{2}$. To this end we
note that $\langle t | n \rangle = \beta_{n}(t)$ implies
\begin{equation}
\sum_{n}\,| n \rangle\,\langle n | t \rangle
=
| t \rangle
=
\sum_{n}\,\beta^{*}_{n}(t)\,| n \rangle.
\end{equation}
With formulas (\ref{eq: eigenstates of R}) for $| n \rangle$
and (\ref{eq: beta definition}) for $\beta_{n}(t)$, the summation 
may be performed and we find
\begin{eqnarray}
&&
| t \rangle = O(t) | 0 \rangle
\nonumber\\
&&
O(t) = N(t)\,\exp{\left(\,-\omega(t)\,L_{+}\,\right)}
\nonumber\\
&&
N(t) = \left[\Gamma(2\,r_{0})\right]^{1/2}\,
\left(
\frac{\omega(t)+1}{2}
\right)^{2\,r_{0}}
\nonumber\\
&&
\omega(t) = \frac{a+\mathrm{i}\,t}{a-\mathrm{i}\,t}
= e\,^{\mathrm{i}\,\theta}
\quad \text{with}~ t = a \tan{\theta/2}. 	
\end{eqnarray}
Thus
\begin{eqnarray}
&&
F_{2}\,\sim\,G_{2}\,\sim\,
\langle 0| O^{\dagger}(t_{1})\,O(t_{2}) |0 \rangle
\nonumber\\
&&
F_{3}\,\sim\,G_{3}\,\sim\,
\langle 0| O^{\dagger}(t_{1})\,B(t)\,O(t_{2}) |0 \rangle.
\end{eqnarray}
We conclude that the averaging state is the $R$ ``vacuum''
$|0 \rangle$ and the operators are $O(t)$ and $O^{\dagger}(t)$.
Note that as anticipated the averaging state is not conformally 
invariant. Also the operators $O$ and $O^{\dagger}$ do not
respond to conformal transformation in the expected way;
they are not primaries.
But these ``defects'' conspire to validate the dAFF realization of
the $SO(2,1)$ generators through the ``t''-derivation, Eq.(\ref{eq: t representation}).
For example, selecting $D$, we form
\begin{eqnarray}
\label{eq: D on state t}
&&
D\,|t \rangle 
= 
\left(
\frac{L_{+}-L_{-}}{2\mathrm{i}}
\right)\,|t \rangle
\nonumber\\
&=&
\left(
\frac{L_{+}-L_{-}}{2\mathrm{i}}
\right)N(t)\,\exp{\left(\,-\omega(t)\,L_{+}\,\right)}|0\rangle
\nonumber\\
&=&
\frac{\mathrm{i}}{2}N(t)
\left(
\frac{d}{d\omega}
e^{-\omega(t)\,L_{+}}
+
\left[
L_{-},e^{-\omega(t)\,L_{+}}
\right]
\right)| 0 \rangle
\end{eqnarray}
The commutator in (\ref{eq: D on state t})
gives
$e^{-\omega(t)\,L_{+}}\,(-2\omega\,R+\omega^{2}\,L_{+})$.
Thus
\begin{eqnarray}
&&
D\,|t \rangle
\nonumber\\ 
&=&
\frac{\mathrm{i}}{2}\,N
\left(
-2\omega r_{0} + (1-\omega^2)\frac{d}{d\omega}
\right)	\frac{1}{N}|t\rangle
\nonumber\\
&=&
-\mathrm{i}\omega r_{0}|t\rangle
+
\frac{\mathrm{i}}{2}(1-\omega^2)
\left(
-\frac{1}{N}\frac{d}{d\omega} N
+
\frac{d}{d\omega}
\right)|t\rangle
\nonumber\\
&=&
-\mathrm{i}
\left(
r_{0}+t\frac{d}{d\,t}
\right)|t\rangle
\end{eqnarray}
in agreement with (\ref{eq: t representation}).
In the last line we used the chain rule 
$t\frac{d}{d\,t} = \frac{1}{2}(1-\omega^2)\frac{d}{d\omega}$.
Similar arguments confirm (\ref{eq: t representation}) for
$H$ and $K$. In a sense $e^{-\omega(t)\,L_{+}}$, when 
acting on $| 0 \rangle$, behaves as a primary operator with
dimension $r_{0}$.

We now demonstrate how conformal constraints arise, even
though the averaging state is not invariant and the operators
do not transform simply.

Consider the expectation value of the commutator with $Q$:
$\langle 0| \left[Q,O^{\dagger}(t_{1})O(t_{2})\right]|0\rangle$,
where $Q$ is any conformal generator. The following equality
holds.
\begin{eqnarray}
\label{eq: conformal variation}
&&
\langle 0| Q O^{\dagger}(t_{1})O(t_{2})|0\rangle
-
\langle 0|O^{\dagger}(t_{1})O(t_{2}) Q|0\rangle
=
\nonumber\\
&&
\begin{split}
&\langle 0| \left[Q,O^{\dagger}(t_{1})\right]O(t_{2})|0\rangle	
\,+
\langle 0| O^{\dagger}(t_{1})\left[Q,O(t_{2})\right]|0\rangle	
\end{split}
\nonumber\\
\end{eqnarray}
When the vacuum is invariant the left side vanishes, because an invariant 
vacuum is annihilated by $Q$. The right side involves the variations
of $O^{\dagger}(t_{1})$ and $O(t_{2})$. Thus one would conclude that 
the conformal variation of the correlation function vanishes.
For us neither is true. The averaging state is not annihilated
by $Q$, which fails to transform $O^{\dagger}(t_{1})$ and
$O(t_{2})$ properly. But the two defects cancel against each other,
thereby establishing the conventional result. This may also be seen by moving
in (\ref{eq: conformal variation}) the left side to the right and canceling
it against the same terms on the right. This leaves the obvious identity
\begin{equation}
0
=
\langle 0|O^{\dagger}(t_{1}) Q O(t_{2})|0\rangle
-
\langle 0|O^{\dagger}(t_{1}) Q O(t_{2})|0\rangle.	
\end{equation}
In order to obtain the invariance constraint using our
CFT$_{1}$ results, we let $Q$ act on the left bra
in the first term, and on the right ket in the second.
With (\ref{eq: t representation}) this produces the 
invariance constraint.

The state 
$|t\rangle = N(t)\,e^{-\omega(t)\,L_{+}}|0\rangle$ is 
like a coherent state, but not quite: it is not an eigenstate
of $L_{-}$ but of $L_{-} + \omega R$.
\begin{equation}
(L_{-} + \omega R)|t\rangle
=
-r_{0}\omega|t\rangle	
\end{equation}

[A conventional coherent state may also be constructed~\cite{Barut}.
\begin{eqnarray}
|\lambda\rangle
&\equiv&
\left[
\Gamma(2 r_{0})
\right]^{1/2}\,
\sum_{n}\,
\frac{\lambda^n}{\left[n! \Gamma(2 r_{0}+n)\right]^{1/2}}
|n\rangle
\nonumber\\
&=&
\Gamma(2 r_{0})
\sum_{n}\,
\frac{\lambda^n}{n! \Gamma(2 r_{0}+n)}
(L_{+})^{n}\,|0\rangle
\nonumber\\
&&
\text{with}~
L_{-}|\lambda\rangle
=
\lambda|\lambda\rangle
\end{eqnarray}
We have used (\ref{eq: eigenstates of R}).
The sum may be performed, but the result yielding a modified
Bessel function with argument $2\sqrt{\lambda\,L_{+}}$ is
not illuminating.] 


An aspect of our construction is noteworthy.
Consider the state $|t\rangle$ at $t=0$, where
$\omega = 1$, and form
\begin{equation}
\label{eq: psi definition}
|\Psi \rangle = e^{-H\,a}|t = 0 \rangle
=
e^{-H\,a}e^{-L_{+}}|0 \rangle.	
\end{equation}
In fact $| \Psi \rangle$ is proportional to the $R$ ``vacuum'' $|0 \rangle$. 
To prove this, act on (\ref{eq: psi definition}) with $R$ and use
\begin{eqnarray}
\label{eq: R commutation}
&&
R\,e^{-H\,a} = e^{-H\,a}
\left(
\frac{K}{2a}+\mathrm{i}\,D
\right)
\nonumber\\
&&
\left(
\frac{K}{2a}+\mathrm{i}\,D
\right)	e^{-L_{+}}
=
e^{-L_{+}}
\left(
R - \frac{1}{4}L_{-}
\right).
\end{eqnarray}
It follows that
\begin{eqnarray}
\label{eq: R on Psi r0}
R	|\Psi \rangle
&=&
e^{-H\,a}e^{-L_{+}}
\left(
R - \frac{1}{4}L_{-}
\right)|0 \rangle	
=
r_{0}e^{-H\,a}e^{-L_{+}}|0 \rangle
\nonumber\\
&=&
r_{0}|\Psi \rangle,
\end{eqnarray}
which establishes that $|\Psi \rangle$
is proportional to $|0\rangle$. In the present
context, this is an example of the operator-state
correspondence:~\cite{Son} the operator $e\,^{-L_{+}}$ with effective
scale dimension $r_{0}$ corresponds to the eigenstate of $R$
with lowest eigenvalue $r_{0}$.

\section{Operator-state correspondence with
neither an invariant vacuum nor a primary operator}
\label{sec: Realizing an operator-state correspondence with
neither an invariant vacuum nor a primary operator}

In dimensions $d \geq 2$, CFT is a quantum field theory and
one usually assumes that a normalized and invariant vacuum state
exists. (This is also true in second quantized quantum mechanics.)
Normal ordering ensures that group generators annihilate the vacuum.
In other words for a field theory we are dealing with a Fock space built
on an empty no-particle vacuum. However quantum mechanics resides in a Hilbert
space, which is a fixed number subspace of the Fock space. This prevents us
from finding a normalized $SO(2,1)$ vacuum state $|\Omega\rangle$ that satisfies
\begin{equation}
\label{eq: annihilation gerenators}
H|\Omega\rangle
=
K|\Omega\rangle
=
D|\Omega\rangle
=
0.	
\end{equation}
A simple way to
see that (\ref{eq: annihilation gerenators}) cannot
be satisfied is by applying the Casimir defined in 
Eq.(\ref{eq: Casimir}): 
$\mathcal{C}|\Omega\rangle = r_{0}(r_{0}-1)|\Omega\rangle \neq 0$
generically. 

Now the definition of a primary field $\mathcal{O}_{\Delta}(t)$
with scaling dimension $\Delta$ is given by the commutation relations
\begin{eqnarray}
\label{eq: primary op}
&&
\mathrm{i}
\left[
H, \mathcal{O}_{\Delta}(0)
\right]
=
\dot{\mathcal{O}}_{\Delta}(0)
\nonumber\\
&&
\mathrm{i}
\left[
D, \mathcal{O}_{\Delta}(0)
\right]
=	
\Delta\,\mathcal{O}_{\Delta}(0)
\nonumber\\
&&
\mathrm{i}
\left[
K, \mathcal{O}_{\Delta}(0)
\right]
=	
0\;,
\end{eqnarray}
where the dot denotes derivative with respect to time.
It would follow from Eqs. (\ref{eq: annihilation gerenators})
and (\ref{eq: primary op}) that 
\begin{equation}
\left(\frac{K}{2\,a}+\mathrm{i}\,D\right)
\mathcal{O}_{\Delta}(0)\,|\Omega\rangle
=
\Delta\,\mathcal{O}_{\Delta}(0)\,|\Omega\rangle.	
\label{eq:op-state}
\end{equation}
We show in the Appendix that there is an operator
$\mathcal{O}_{\Delta}(0)$ and a non-normalizable ``state'' $|\Omega\rangle$ that together
conspire to satisfy Eq.~(\ref{eq:op-state}), even though
$\mathcal{O}_{\Delta}(0)$ fails to satisfy (\ref{eq: primary op}), 
and $|\Omega\rangle$ is annihilated only by $H$ and not by $D$ and $K$.
This allows us to define the state $|{\cal O}_\Delta\rangle\equiv
\mathcal{O}_{\Delta}(0)\,|\Omega\rangle$ and we can proceed
as usual to obtain the correlation functions of the CFT$_1$. To do so,
one deduces with (\ref{eq:op-state}) that
\begin{eqnarray}
\label{eq: R on Psi}
R\,e^{-Ha}|{\cal O}_\Delta\rangle 
&=& 
e^{-Ha}\left(\frac{K}{2\,a}+\mathrm{i}\,D\right)
|{\cal O}_\Delta\rangle
\nonumber\\
&=&
\Delta\, e^{-Ha}\,|{\cal O}_\Delta\rangle
\;,
\end{eqnarray}
from which it follows that
\begin{equation}
e^{-Ha}\,|{\cal O}_{r_0}\rangle\propto |0 \rangle
\;.
\end{equation}

We now show that correlation functions 
of fields computed in the state $|\Omega\rangle$
have the same scaling behavior as
the matrix elements~(\ref{eq: 2-pt function dFF})
and~(\ref{eq: 3-pt function dFF}) computed in the t-representation
of dAFF. Considering explicitly the case of the $2$-point correlation
function, we can define
\begin{eqnarray}
\label{eq: 2-pt-correlation-op-state}
&&
G_{2}(t_{1},t_{2})
=
\langle \Omega|\,
\mathcal{O}_{r_{0}}(t_{1})\,
\mathcal{O}_{r_{0}}(t_{2})
\,|\Omega\rangle	
\nonumber\\
&=&
\langle \Omega|\,\mathcal{O}_{r_{0}}(0)
\,
e\,^{-\mathrm{i}(t_{1}-t_{2})H}\,
\mathcal{O}_{r_{0}}(0)\,
\,|\Omega\rangle
\nonumber\\
&=&
\langle \Omega|\,\mathcal{O}_{r_{0}}(0)
\,e\,^{-Ha}\,e\,^{\left[2\,a-\mathrm{i}(t_{1}-t_{2})\right]H}\,
e\,^{-Ha}\,\mathcal{O}_{r_{0}}(0)\,
\,|\Omega\rangle
\nonumber\\
&=&
\langle 0|\,
e\,^{\left[2\,a-\mathrm{i}(t_{1}-t_{2})\right]H}
\,|0 \rangle
\end{eqnarray}
in which the time translation invariance of $G_{2}(t_{1},t_{2})= G_{2}(t_{1}-t_{2}) \equiv G_{2}(t)$ is made explicit 
in the second line of~(\ref{eq: 2-pt-correlation-op-state}) due to $H$ annihilating $|\Omega\rangle$.
It is then straightforward to show,
by differentiation of the last line of~(\ref{eq: 2-pt-correlation-op-state}) with respect to time,
that $G_{2}(t)$ satisfies
\begin{equation}
\label{eq: 2pt diff equation}
\left(
t\,\frac{\partial}{\partial\,t}
+
2\,r_{0}
\right)
G_{2}(t)
=
0\;.	
\end{equation}
Solution of~(\ref{eq: 2pt diff equation}) yields 
\begin{equation}
\label{eq: 2-pt function scaling op-st}
G_{2}(t) \sim |t|^{-2\,r_0}
\;.	
\end{equation}
As anticipated, the $2$-point correlation function
defined in~(\ref{eq: 2-pt-correlation-op-state}) displays the same scaling behavior
as~(\ref{eq: 2-pt function dFF}). Similarly, the three point function
\begin{equation}
\label{eq: 3-pt-correlation-op-state}
G_{3}(t;\,t_{2},t_{1})
=
\langle \Omega|
\mathcal{O}_{r_{0}}(t_{2})\,
B(t)\,
\mathcal{O}_{r_{0}}(t_{1})|
\Omega\rangle\;	
\end{equation}
constructed with a primary operator $B(t)$ satisfying~(\ref{eq: primary op})
with scaling dimension $\delta$, has the form
\begin{equation}
\label{eq: 3-pt function scaling op-st}
G_{3}(t;\,t_{2},t_{1})
\sim
\frac{1}
{
|t-t_{1}|^{\delta}
\,
|t_{2}-t|^{\delta}
\,
|t_{1}-t_{2}|^{-\delta + 2r_{0}}
}\;.	
\end{equation}
This scaling behavior is the same as~(\ref{eq: 3-pt function dFF}). 
From the form of the correlation functions~(\ref{eq: 2-pt function scaling op-st})
and~(\ref{eq: 3-pt function scaling op-st}) one learns that,
despite the fact that neither the state $|\Omega\rangle$ is 
annihilated by all the generators of the conformal group nor the
operator $\mathcal{O}_{r_{0}}$ is primary, still the 
combination $\mathcal{O}_{r_{0}}|\Omega\rangle$ makes the correlation
functions to have the scaling behaviors that are expected as if they were
built out of a fully invariant vacuum state and a primary operator.

\section{Conclusion}
\label{sec: Conclusion}

Motivated by the conjectured AdS$_{d+1}$/CFT$_{d}$
correspondence, and by the least understood but relevant
case of the AdS$_{2}$/CFT$_{1}$ correspondence, we have studied
the properties of conformally invariant quantum mechanics. 
Given the fact that in the CFT$_{1}$ one has to deal with a Hilbert space
(instead of a Fock space when $d \geq 2$), we find that the
usual operator-state correspondence needs to be modified. However,
such a modification still allows one to build up correlation functions
that behave as if they were constructed out of a fully invariant vacuum (i.e., annihilated by all the group generators) 
and primary operators carrying well defined scale dimension.
This result is established with only implicit reference to a dynamical CFT$_{1}$ model. 
Rather, our derivation exploits the $SO(2,1)$ group structure. However, we also have in hand an explicit dynamical model for the CFT side of the correspondence, so that many features can be evaluated in a controllable way. It would be interesting to construct the corresponding dual AdS2 field theory.
 We leave this question to a future investigation.


\section*{
Acknowledgments
         }

We acknowledge useful discussions with Allan Adams, Sean Hartnoll and
Andy Strominger. This research is supported by DOE grants
DEF-06ER46316 (CC), -91ER40676 (S-Y P) and DE-FG02-05ER41360 (RJ).

\appendix
\section*{Appendix}
\label{sec:appendix}

Here we clarify some of the steps related to the realization 
of the operator-state correspondence in the CFT$_{1}$ model
defined in~(\ref{eq: dFF model}). Considering first Eq.~(\ref{eq:op-state})
for $\langle q|\mathcal{O}_{\Delta}\rangle$:
\begin{equation}
\left[
\frac{q^{2}}{4\,a}
-
\frac{1}{4}
\left(
\frac{d}{d\,q}\,q
+
q\,\frac{d}{d\,q}
\right)
\right]
\,
\langle q|\mathcal{O}_{\Delta}\rangle
=
\Delta\,\langle q|\mathcal{O}_{\Delta}\rangle\;,	
\end{equation}
which gives
\begin{equation}
\label{eq: state O in q representation}
\langle q|\mathcal{O}_{\Delta}\rangle
=
e\,^{\frac{q^{2}}{4a}}\,
q^{-\frac{1}{2}-2\,\Delta}\;.	
\end{equation}
Now $\langle q|\Omega\rangle$ is found by using $H|\Omega\rangle = 0$ and solving
\begin{equation}
\langle q|H|\Omega\rangle
=
\frac{1}{2}
\left(
-\frac{d\,^2}{d\,q^2}+\frac{g}{q^2}
\right)\langle q|\Omega\rangle
=
0\;.	
\end{equation}
The solution is non-normalizable.
\begin{equation}
\label{eq: Omega}
\langle q|\Omega\rangle \sim q^{r_0}	
\end{equation}
Now from the definition
\begin{equation}
|\mathcal{O}_{\Delta}\rangle	
=
\mathcal{O}_{\Delta}(0)\,|\Omega\rangle\;
\end{equation}
and assuming that the operator $\mathcal{O}_{\Delta}(0)$ is diagonal in the $q$ representation, i.e., 
\begin{equation}
\langle q | \mathcal{O}_{\Delta}(0) | q' \rangle
=
V(q)\,\delta(q-q')\;,	
\end{equation} 
one gets, by~(\ref{eq: state O in q representation})
and~(\ref{eq: Omega})  
\begin{equation}
V(q)
=
e\,^{\frac{q^2}{4a}}\,q^{-\left( \frac{1}{2} + 2\Delta + r_0 \right)}\;.
\end{equation}
The action of $K$ and $D$ on the state $|\Omega\rangle$ can be straightforwardly computed
\begin{eqnarray}
\label{eq: K and D on Omega}
&&
\langle q | K | \Omega\rangle
=
\frac{q^2}{2}\,\langle q|\Omega\rangle	
\nonumber\\
&&
\begin{split}
\langle q | D | \Omega\rangle
&=
\frac{\mathrm{i}}{4}\,
\left(
\frac{d}{d\,q}\,q
+
q\,\frac{d}{d\,q}
\right)\,\langle q|\Omega\rangle
\\
&=
\frac{\mathrm{i}}{4}\,
(2\,r_{0} + 1)\,\langle q|\Omega\rangle&\,
\end{split}	
\end{eqnarray}
Eqs.~(\ref{eq: K and D on Omega})
show explicitly that $|\Omega\rangle$ is not annihilated either by $K$ or $D$.
Moreover, the commutation relations
\begin{eqnarray}
\label{eq: K and D commutators with O}
&&
\mathrm{i}\,
\left[
K, \mathcal{O}_{\Delta}(0)
\right]	= 0
\nonumber\\
&&
\mathrm{i}\,
\left[
D, \mathcal{O}_{\Delta}(0)
\right]	
=
\mathcal{O}_{\Delta}(0)
\left[
\left( \frac{1}{4} + \Delta + \frac{r_0}{2}  \right)
-
\frac{K}{2\,a}
\right]\;	
\end{eqnarray}
make manifest that $\mathcal{O}_{\Delta}(0)$ is not 
a primary operator satisfying~(\ref{eq: primary op}).
Nevertheless, (\ref{eq: K and D on Omega}) and
(\ref{eq: K and D commutators with O}) combined yield (\ref{eq:op-state}).

\end{document}